# Plasmonic nanolaser for intracavity spectroscopy and sensorics

P. Melentyev,[1] A. Kalmykov,[1,2] A. Gritchenko,[1,3] A. Afanasiev,[1] V. Balykin,[1]
A. Baburin,[4] E. Ryzhova,[4] I. Filippov,[4] I. Rodionov,[4]
I. Nechepurenko,[6] A. Dorofeenko,[3,5,6] I. Ryzhikov,[3,5] A. Vinogradov,[3,5,6]
A. Zyablovsky,[6] E. Andrianov,[3,6] and A.A. Lisyansky[7,8]

[1]*Institute of Spectroscopy RAS, Troitsk, Moscow, 142190, Russia*
[2]*National Research University, Higher School of Economics, Moscow, 101000, Russia*
[3]*Moscow Institute of Physics and Technology, Dolgoprudny, 141700, Russia*
[4]*Bauman Moscow State Technical University, Moscow, 105005, Russia*
[5]*Institute for Theoretical and Applied Electromagnetics RAS, Moscow, 125412, Russia*
[6]*Dukhov Research Institute of Automatics, Moscow, 127055, Russia*
[7]*Physics Department, Queens College of the City University of New York, Queens, New York, 11367, USA*
[8]*The Graduate Center of the City University of New York, New York, New York 10016, USA*

Corresponding authors' emails: melentiev@isan.troitsk.ru (P.M.), balykin@isan.troitsk.ru (V.B.), irodionov@bmstu.ru (I.R.), alexander.lisyansky@qc.cuny.edu (A.A.L.).

We demonstrate intracavity plasmonic laser spectroscopy using a plasmonic laser created from a periodically-perforated silver film with a liquid gain medium. An active zone of the laser is formed by a highly elongated spot of pumping. This results in an 80x decrease in the threshold pumping level; in a significantly more efficient diffusive mixing of dye molecules, which substantially suppresses the effect of their bleaching; and in the ability to reduce the volume of the gain medium to as little as 400 nL. We use this design for a stable plasmonic laser in multiple measurements and demonstrate that it is highly effective as a spaser spectroscopy sensor for intracavity detection of an absorptive dye at 70 ppb. This work provides an opportunity to develop applications of intracavity plasmonic laser spectroscopy in biological label detection and other fields.

In laser physics, intracavity laser spectroscopy methods are known to have an extremely high sensitivity.[1] In these methods, an analyte is placed in the laser cavity. This analyte should absorb at the lasing wavelength. As a consequence, the analyte is detected by changing the lasing parameters, such as the threshold, lasing intensity and wavelength. Usually, it manifests in an appearance of dips in the lasing spectrum.

The dynamics of a plasmonic laser, or spaser, share many characteristics with that of a traditional laser.[2-7] Plasmonic laser intracavity spectroscopy was first proposed theoretically.[8-10] In particular, it has been shown that, similar to conventional intracavity laser spectroscopy, the use of plasmonic lasers can lead to an increase in the efficiency of plasmonic sensors by up to two orders of magnitude.[8]

Different successful experimental approaches have been recently reported.[11-13] In these studies, a spaser based on quantum wells has been sunk into the surrounding medium; an addition of an analyte to that medium results in the modification of the spaser generation frequency[11,12] or intensity.[13] Here, as opposed to conventional intracavity spectroscopy, an addition of an analyte results in an increase of gain and, as a consequence, the laser intensity is boosted.

In this study, we deal with the conventional approach in the sensor design, in which an addition of an analyte is assumed to decrease the intensity of radiation. To begin with, we design a *planar plasmonic laser*, extending the idea and implementation of plasmonic lasers reported elsewhere.[14-23] We use a liquid gain medium and a perforated metallic film. The use of this "flow-through" liquid gain enables the direct introduction of an analyte into the active region that warrants both the maximal sensitivity and multiple usage of the sensor. We expect that the subwavelength scale of the plasmonic cavity and the employment of microfluidics make plasmonic intracavity spectroscopy suitable for lab-on-chip technology.

It seems that for the design of a plasmonic laser the greater gain the better. However, it has been recently shown that dye molecules are optimal for plasmonic lasers because usage of quantum wells with greater gain results in energy loss via spontaneous emission.[24] Moreover, the dye molecules are capable of nearly 100% yield and allow for the creation of an optically homogeneous medium with high gain.[25] The main shortcoming of this type of active medium is photobleaching, in which a dye molecule is unable to continue fluorescence after approximately $10^5-10^6$ photon emissions.[26] In other words, in the CW regime, a laser using dye in a polymer matrix lives for only 1 ms.

In this paper, we describe an implementation of a pulsed pumped plasmonic laser that uses a *solution* of R101 dye in dimethyl sulfoxide (DMSO) as the gain medium. The pulse duration and energy should be small enough so that photobleaching does not occur. The pumping should be sufficient to saturate the excitation of dye molecules. Experiment shows that the required pump power is about 1 MW/cm$^2$. Such pumping causes bleaching of dye molecules in 100 µs.[26] Thus, the pump pulse duration should be shorter than this value.

The use of a liquid gain medium offers a number of advantages over a solid gain medium by fixing the bleaching problem. In a DMSO solution, dye molecules diffuse for a distance of $l \approx \sqrt{2Dt}$ with the diffusion coefficient $D \approx 0.1 \, \text{cm}^2/\text{s}$. In our experimental setup, the size of the pumped region is 1 mm and the diffusion length is 1 mm. The exposure time of 0.5 s determines the repeating frequency of 2 Hz. This ensures that each pulse interacts with new dye molecules. Therefore, photobleaching is highly suppressed and a temporarily stable gain medium is formed.

Fig. 1a shows the setup used in this study. It is a planar waveguide made of a liquid DMSO layer deposited on a plasmonic crystal. The plasmonic crystal is formed by an array of nanoholes created in a 100-nm-thin silver film deposited onto a quartz substrate. The optimal parameters of the plasmonic crystal have been numerically calculated. To compensate the uncertainty in calculated values of the material parameters, we fabricated and studied structures with nanoholes of the varying pitch Λ (545, 555, 565, 575, and 585 nm) and the diameter (150, 175, and 200 nm).

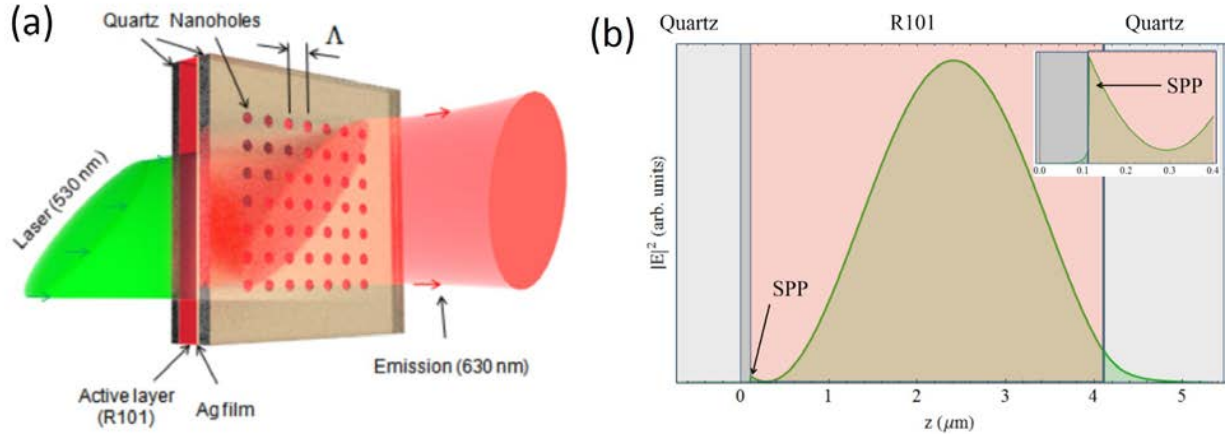

FIG. 1. Planar plasmonic laser. (a) The laser apparatus using a perforated silver film. (b) The spatial distribution of the electric field intensity of the TM polarized mode in the quartz/silver (100 nm)/DMSO (1000 nm)/quartz system. The inset shows the electric field near the silver/DMSO interface.

The nanohole arrays in the silver films are fabricated by using electron-beam lithography and dry etching. As shown in Fig. 2, the array of holes is patterned on a $100 \pm 5$ nm thick silver film, which is deposited on a UV-grade quartz substrate with a roughness under 1 nm. The silver film is deposited by using multistep electron-beam evaporation with a base pressure of $5 \times 10^{-8}$ torr; then it is capped by a 10-nm-layer of $SiO_2$. The process for silver film multistep electron-beam evaporation on silver films is developed as the part of the detailed study of silver film formation mechanisms.[27] The quality of the silver films is monitored by scanning electron microscopy (SEM) and electron backscatter diffraction (EBSD) yielding an average film grain size more than 1 μm, as well as by stylus profilometry, yielding an rms roughness less than 1.2 nm, and by spectroscopic ellipsometry, which results in optical properties similar to those reported in Ref. [28]. The use of the $SiO_2$-film solves two issues: it suppresses the silver degradation and prevents the quenching of excited dye molecules. The quenching leads to an undesirable heating of the silver film that, in turn, leads to sample destruction at pumping intensities above 2 MW/cm$^2$.

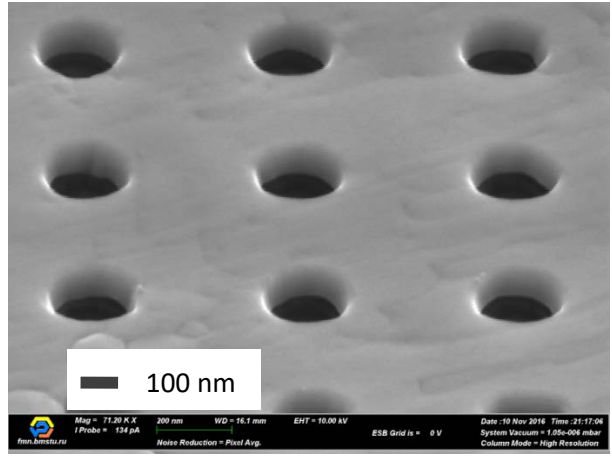

FIG. 2. SEM images of an array of 200 nm nanoholes with a pitch of 565 nm fabricated in a single silver crystal.

Since the losses in silver are large, we deal with hybrid modes. In principle, the optimal lasing parameters are obtained when the bulk of the mode energy is contained in the DMSO layer with dye molecules. In this study, we experimentally determine that for lasing, the minimal possible gain layer thickness should be 4 μm. The calculated field distribution in the *TM* mode is shown in Fig. 1b. Hybridization between the photonic and plasmonic modes appears at a maximum at the DMSO/metal interface. As a result, the interaction of the modes with the nanoholes is high enough to induce the formation of a Bloch wave with the attenuation coefficient[29] as low as 1 cm$^{-1}$.

The gain medium is formed by the solution of the R101 dye in DMSO with the concentration $6 \times 10^{18}$ cm$^{-3}$. This system provides a material gain near 75 cm$^{-1}$ via pulsed optical pumping at the wavelength of 530 nm. To decrease the threshold pumping, we have chosen a specific geometry of the pumping beam, which selectively excites some modes. In particular, the pumping region takes the form of an extended ellipse with the minor and major axes equal to 1.6 mm and 20 μm, respectively. This geometry leads to a decrease in the threshold pumping by the factor of 80, and consequently, to an increase in the lifetime of the gain medium, compared to the case of a circular pumping spot with a diameter equal to the major axis of the ellipse. The major axis is aligned along the 1.4 mm diagonal of the nanohole array. Therefore, 0.1 mm of the pumped region lies outside of the nanohole array on each side. The peak pumping strength is varied up to the value of 5 MW/cm$^2$, significantly exceeding the saturation intensity of the R101 optical transition. Furthermore, the dye diffusion occurs more effectively owing to the narrow pumped region; hence, the repeating frequency can be increased up to 40 Hz with a pulse duration of 10 ns. The volume of the gain medium in the pumped region is by four orders of magnitude smaller than the volume of the total gain medium. This leads to a long-lived stable generation regime that can be sustained for several weeks with the use of only 400 nL of the gain medium.

In order to characterize the system, the spectrum of dye luminescence is initially measured for the system without nanoholes. Radiation is brought out through a slit in the metal film with the width of about 2 µm. In Fig. 3a, one can see the change in the luminescence spectrum depending on the pumping level. This change demonstrates a threshold behavior. At low pumping, below the threshold, the luminescence spectrum of the sample coincided with that of a single R101 dye molecule. It has a maximum at 620 nm and the spectral width at half-maximum being equal to 40 nm. Above the threshold, at high pumping (> 2.5 MW/cm$^2$), a narrower, 12 nm, spectral line centered at 618 nm emerges. This line and the presence of the threshold is characteristic of the multimode amplified spontaneous emission.[24]

Next, we consider the situation when the elongated pumping spot covers the perforated area of silver in the (1,1)-direction. There are boundaries between the plane surface and the perforated area which form a resonator. The modes of this resonator are Bloch waves. The lasing Bloch modes are leaky, which results in the output signal at some angle $\theta_l$ to the sample normal.

Though the reflectance coefficients of the lasing Bloch waves from the resonator boundaries are small, this is enough to form lasing. The smallest lasing threshold occurs for the sample with 175 nm nanoholes and a pitch of 565 nm, and the spectrum of the plasmonic laser is measured in the leaky lobe at the angle $\theta_l = 5°$.

The results for two values of pumping are shown in Fig 3b. At intensities below 500 kW/cm$^2$, the spectral contour is very similar to that of usual R101 dye luminescence. However, at higher (above 500 kW/cm$^2$) pumping intensities, the luminescence contour changes dramatically, and a very narrow spectral line with the half-width of 1.7 nm emerges at 628 nm. Such a narrow line width indicates that we deal not with amplified spontaneous emission (ASE) but with lasing.[24]

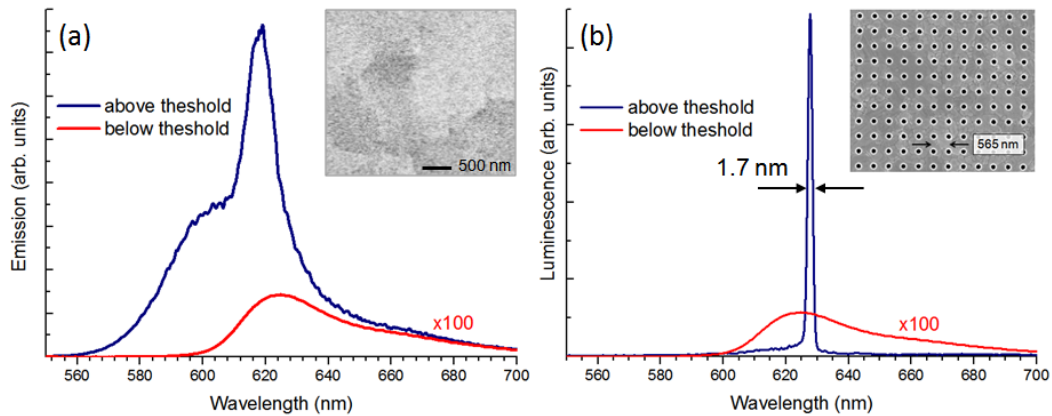

FIG. 3. Measured radiation spectra of different samples. (a) The spectrum of dye luminescence in the quartz/silver/DMSO/quartz system without nanoholes, with radiation output through a defect in the form of the 5-µm-wide slit passing through the film. (b)The luminescence spectrum of the plasmonic laser. The insets show electronic microscope images of the samples.

Fig. 4 shows the dependence of the lasing intensity and the spectral width on the pumping strength obtained in the experiment. These measurements are carried out by imaging the microscope focal plane onto the CCD camera by means of a Bertrand lens.[30] We observe a threshold behavior of plasmonic laser radiation with the threshold pumping value of 500 kW/cm$^2$ corresponding to a gain coefficient ~40 cm$^{-1}$.

The insets in Fig. 4 show the dependence of the angular radiation distribution of the plasmonic laser for pumping intensities below and above the threshold. Above the plasmonic lasing threshold, directional radiation is observed. The radiation pattern is extended in the direction perpendicular to the extension direction of the pumping region. The minimal directivity is ~1.3°, however, in the orthogonal direction, it is 4.1°.

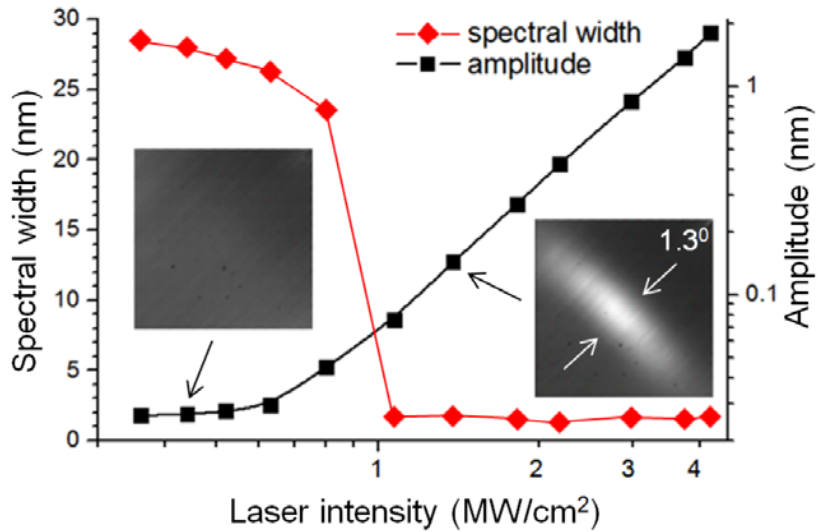

FIG. 4. Generation intensity and line width dependence of the plasmonic laser. Insets show the angular distribution of the intensity in the radiation pattern for pumping intensities below and above the lasing threshold.

Let us demonstrate the possibility of applying the demonstrated system in sensing and intracavity spectroscopy.[8-10] We show that a presence of a small concentration of an absorptive analyte inside the lasing cavity leads to a substantial suppression of lasing. The change in the output intensity is treated as a sensor response. Using a liquid gain medium allows for the addition of analyte directly to the gain solution. In this case, high lasing mode intersection with the gain medium guarantees the same intersection with the analyte.

We also demonstrate the possibility of sensing applications with the plasmonic laser as implemented with the apparatus shown in the inset of Fig. 5. Via a microfluidic, input we introduce an analyte to the R101 dye solution, cyanine-5 dye, which has a broad absorption band around the lasing spectral region. Solutions with three concentrations, 7, 0.7, and 0.07 ppm, have been used. Before their injection, the gain medium without cianine-5 dye has been removed from the lasing cavity. We use the solution volume of 800 nL, twice than that of the dye solution above the nanohole array. Therefore, the measurement at each analyte concentration requires an extremely small probe volume, which is a principal feature of our sensor.

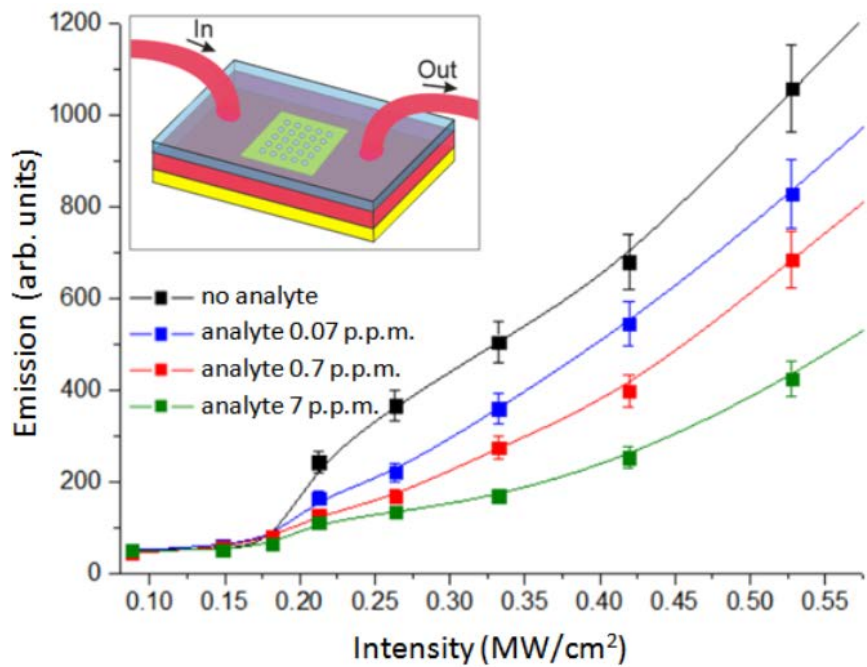

FIG. 5. Lasing intensity vs. pumping rate at different cyanine-5 dye concentrations in the lasing cavity. The sketch of the experimental setup is shown in the inset.

The resulting lasing curves are shown in Fig. 5. In addition to the lasing suppression by the absorptive dye, we detect a minimal concentration of 70 ppb, thereby proving the ultra-high detection sensitivity of our novel scheme. This concentration corresponds to the theoretical estimates of the detection limit based on the dye absorption cross-section.

We implement and study the plasmonic laser generation based on hybrid modes of a plasmonic crystal and a photonic waveguide interacting with the gain medium of R101 dye solution. The laser cavity is formed by a nanohole array in a silver film. The threshold behavior of the radiation characteristics with an increase in the pump power is demonstrated along with the high directivity above the threshold. The plasmonic laser cavity is characterized by an ultra-low loss (40 cm$^{-1}$), owing to the use of ultra-high-quality silver film and the hybrid plasmonic-photonic mode. Lasing at 628 nm, with a line width of 1.7 nm and directivity of 1.3° is observed. Based on the implemented scheme, we also demonstrate a possibility of intracavity plasmonic laser spectroscopy using an analyte concentration of 70 ppb.

We would like to thank Yu. Lozovik, Yu. Vainer, R. Gordon, A. Schokker, F. Koenderink, and A. Kalatskiy for helpful discussions. The research was supported by the Advanced Research Foundation (Contract No. 7/004/2013-2018) and Russian Foundation for Basic Research (Grant No. 17-02-01093). Samples were made at the BMSTU Nanofabrication Facility (Functional

Micro/Nanosystems, FMNS REC, ID 74300). A.A.L. acknowledges support from the NSF under Grant No. DMR-1312707.